\documentclass[12pt]{article}
\usepackage[utf8]{inputenc}
\usepackage[linesnumbered,ruled,vlined]{algorithm2e}
\usepackage{calrsfs}
\usepackage{graphicx,psfrag,epsf,xcolor}
\usepackage{enumerate}
\usepackage{dsfont}
\usepackage{natbib}
\usepackage{subcaption}
\usepackage{booktabs}
\usepackage{epsfig}
\usepackage{longtable}
\DeclareMathAlphabet{\pazocal}{OMS}{zplm}{m}{n}
\usepackage{amsmath,amsfonts,amsthm,bm} 
\usepackage{setspace}
\newcommand{\blind}{0}

\begin{document}

\def\spacingset#1{\renewcommand{\baselinestretch}%
{#1}\small\normalsize} \spacingset{1}


\if0\blind
{
  \title{\bf Hierarchical Dirichlet Process Mixture of Products of Multinomial Distributions: Applications to Survey Data with Potentially Missing Values}
  \author{Chayut Wongkamthong \hspace{.2cm}\\
    Kasikorn Labs,\\
    Kasikorn Business-Technology Group}
  \maketitle
} \fi

\bigskip
\begin{abstract}
In social science research, understanding latent structures in populations through survey data with categorical responses is a common and important task. Traditional methods like Factor Analysis and Latent Class Analysis have limitations, particularly in handling categorical data and accommodating mixed memberships in latent structures, respectively. Moreover, choosing the number of factors or latent classes is often subjective and can be challenging in the presence of missing values. This study introduces a Hierarchical Dirichlet Process Mixture of Products of Multinomial Distributions (HDPMPM) model, which leverages the flexibility of nonparametric Bayesian methods to address these limitations. The HDPMPM model allows for multiple latent classes within individuals and avoids fixing the number of mixture components at an arbitrary number. Additionally, it incorporates missing data imputation directly into the model's Gibbs sampling process. By applying a truncated stick-breaking representation of the Dirichlet process, we can derive a Gibbs sampling scheme for posterior inference. An application of the HDPMPM model to the 2016 American National Election Study (ANES) data demonstrates its effectiveness in identifying political profiles and handling missing data scenarios, including those that are missing at random (MAR) and missing completely at random (MCAR). The results show that the HDPMPM model successfully recovers dominant profiles and manages complex latent structures in survey data, providing an alternative tool for social science researchers in dealing with categorical data with missing values.
\end{abstract}

\noindent%

\noindent
\textbf{Statement of significance}: This research introduces the Hierarchical Dirichlet Process Mixture of Products of Multinomial Distributions (HDPMPM) model, advancing the methodology for analyzing survey data with categorical responses. By overcoming the limitations of traditional techniques, such as Factor Analysis and Latent Class Analysis, the HDPMPM model introduces a flexible nonparametric Bayesian approach that accommodates multiple latent classes within individuals and a potentially infinite number of mixture components. Importantly, it also incorporates missing data imputation within its Gibbs sampling process. The practical utility of this model is demonstrated using data from the 2016 American National Election Study (ANES), successfully identifying intricate political profiles and addressing various missing data scenarios. This research offers an alternative tool for social science researchers, enabling more comprehensive analysis of complex survey datasets, thereby contributing to a deeper understanding of latent structures in population studies.

\doublespacing
\section{Introduction}
In many social science applications, there is often an interest in studying and analyzing latent structures within certain sub-populations using survey data containing many categorical responses. For example, when studying political ideology and public opinion surveys, the goal is usually to uncover and explain the structure of different political ideologies that may exist in the population \citep{Fleishman1986, Treier2009, gross:2012, Alvarez2017, Silva2018}. The simplest example of such typologies consists of a liberal profile and a conservative profile, although typologies with more than two groups are also common \citep{Fleishman1986, pew2011}. 

To achieve this goal, analysts usually start by assuming that people rely on their basic orientations or principles when evaluating a political idea \citep{feldman:1988}. In contrast to the idea of \citet{zaller1992}, many scholars believe that the orientations of people can vary and may not be unidimensional \citep{americanvoter, Converse2006}. With this assumption, responses to survey questions are conditionally independent, given the latent variables that capture people's ideologies. Then, depending on the form of political configurations they believe people have, analysts rely on some analytical tools to conceptualize this latent structure. Among these, Factor Analysis (FA, \citet{spearman04}) is probably the most popular technique for assessing political ideology due to its close relation to the ordinary linear regression and its early use in this field of study \citep{feldman:1988, Converse2006, LAMERIS2018417}. However, this method treats survey responses as continuous outcomes while responses to survey questions in political polls are usually categorical.


Another popular technique is Latent Class Analysis (LCA, \citet{goodman1974}), which can be used for categorical variables. In this context, LCA assumes that each person is associated with exactly one political typology or profile, and, for all members within each distinct profile, responses to a question are independent and identically distributed as per the distribution associated with that typology. For more discussion on LCA and its application in social science, see \citet{Taylor1983,MCCUTCHEON1985, BREEN_2000} and \citet{Hagenaars2003}. Even though LCA provides a simple tool to cluster people into different political profiles, its complete assignment of each person to one and only one political type is quite restrictive. This assumption forces each individual to fully conform to exactly one political typology, leaving no opportunity for one's ideology to fall in between different political profiles. This also contradicts some scholars' views of political ideology as a spectrum \citep{maddox:1984, feldman:2009}. 

The Grade of Membership (GoM, \citet{woodbury:1978, Erosheva:2007}) model solves this issue by allowing multiple political typologies to coexist within a person. Under its Mixed Membership (MM) framework, each individual is assumed to have their own membership vector, which shows their degree of association with different possible political profiles. To respond to a political question, the person first randomly draws a political profile based on their degree of association in the membership vector. Then, given that profile, the person expresses their opinion according to the distribution associated with that profile. For its applications in political ideology study, see \cite{gross:2012}. Even though this model is intuitive and flexible, it requires the number of mixture components as an input, while this number is unknown most of the time. 

One way to avoid fixing the number of mixture components is to rely on nonparametric models. An important method when dealing with multivariate categorical responses is the Dirichlet process (DP) mixture of products of multinomial distributions (DPMPM, \citet{DPMPM_Kenough}) model. However, the DPMPM model assumes that each person belongs to one particular cluster component and hence does not have the mixed membership property that we are seeking. For examples of using DPMPM in political science, see \citet{Si_Reiter_Hillygus_2015, Meyer2019UsingTD}.

More generally, consider a setting where we have multivariate categorical responses from people in a population. We consider responses to various prompts from the same person as one group of variables. Our goal is to derive a model that takes into account mixtures that represent the latent structure in each group. Specifically, we allow responses from a person to be associated with different mixtures under the MM paradigm. To obtain global or population-level structure, we want the mixture components or parameters to be shared across different groups. For practical purposes, this study assumes that the total number of mixture components is unknown. This type of latent structure usually arises in surveys in various application domains, including social science \citep{gross:2012} and healthcare \citep{Erosheva:2007}. For a more detailed discussion on the MM paradigm and its applications, see \citet{Airoldi2014}.

There are several existing methods that analysts normally use in modeling datasets with such latent structures as discussed previously. However, to the best of our knowledge, these methods either assign each group of variables to the same cluster (total assignment) or require some prior knowledge about the exact number of mixture components, which may not be available in practice. This paper presents a nonparametric Bayesian mixed membership model that suitably handles datasets with such latent structures. The author calls this method a hierarchical Dirichlet process mixture of products of multinomial distributions model, abbreviated as the HDPMPM model. The basic idea of the HDPMPM model is to use Hierarchical Dirichlet Processes (HDP, \citet{teh:2006}) as the prior distribution for the components of the mixture of products of multinomial distributions. In this model, responses from each person are modeled as a mixture of products of multinomial distributions. Each set of mixture components for each person is assumed to be an independent draw from a Dirichlet process with the same global probability measure $G_0$. To make $G_0$ discrete, which will allow the sharing of the mixture components across different people, $G_0$ is drawn from another Dirichlet process with a base probability measure $H$, which is a product of independent flat Dirichlet distributions in this case. 

This paper provides a Gibbs sampling scheme to perform posterior inference of the HDPMPM model using the truncated stick-breaking representation of the DP \citep{Ishwaran2001}. To the best of our knowledge, there are still no studies that provide detailed full conditional distributions for all parameters of the HDP with mixture of products of multinomial distributions using the truncated stick-breaking representation.  This method differs from schemes presented by \citet{teh:2006} as it uses the truncated stick-breaking representation of the DP and does not completely integrate out any membership vectors of any observation groups. This allows us to investigate the degree of alignment of each observation group toward any profiles of interest. We also note that this is a more specialized version of the scheme presented in \citet{ren:2008} where the probability of innovation is 1. 


This study shows an application of the HDPMPM model in analyzing political profiles in a hypothetical population using data from the 2016 American National Election Study (ANES). Moreover, since missing data is a very common problem in survey studies, an imputation method that can naturally be integrated into the Gibbs sampling of the HDPMPM model is provided. The performance of the HDPMPM model with missing data imputation is illustrated by reconstructing some dominant profiles from the same population dataset under both the missing at random (MAR, \citet{littlerubin}) and the missing completely at random (MCAR, \citet{littlerubin}) scenarios. We note that the main goal here is not to thoroughly investigate political ideology and its configurations in this dataset. We discuss basic analyses on this dataset solely to show an application of the HDPMPM model to survey data that may have this complicated latent structure. For a detailed discussion on assessing political ideology from surveys, see \cite{Fleishman1986, feldman:1988, gross:2012}. 


The remainder of this article is organized as follows. Section \ref{section_review} provides a review of some of the existing methods commonly used to analyze categorical data, possibly with the aforementioned latent structure. In section \ref{section_HDPMPM}, the HDPMPM model using the truncated stick-breaking representation of HDP is presented. The section also describes the Gibbs sampling method for posterior inference of the HDPMPM model. Section \ref{section_ANES_appl} provides data description from the 2016 ANES. The use of the HDPMPM model in analyzing political ideology using a hypothetical population derived from the ANES is also illustrated. In section \ref{section_missing_data}, we discuss the use of the HDPMPM model in the presence of missing values. The paper demonstrates the performance of the method in recovering the dominant profiles using the same population dataset under both the MAR and the MCAR scenarios. In section \ref{section_conclusion}, we conclude with general takeaways.

\section{Review of the DPMPM and the GoM Models} \label{section_review}
Before presenting the proposed approach, we first review two closely related mixture models for analyzing multivariate categorical data: (i) the Dirichlet process mixture of products of multinomial distributions (DPMPM, \citet{DPMPM_Kenough}) model, and (ii) the Grade of Membership (GoM, \citet{woodbury:1978, Erosheva:2007}) model. We begin with the notation needed to understand the models.

\subsection{Notation} 
Let $\bm{X} = (\bm{x}_1,\ldots,\bm{x}_n)$ denote the $n \times p$ matrix containing the data for all $n$ observations across $p$ variables. For $i = 1,\ldots,n$, let $\bm{x}_i = (x_{i1},\ldots,x_{ip})$, where each $x_{ij} \in \{1,2,\ldots,D_j\}$ represents the value of variable $j$ out of $D_j$ levels for individual $i$. Also, let $\bm{x}_j = (x_{1j},\ldots,x_{nj})$. Let $\bm{R}$ denote the missing value indicator matrix where, for observation $i$ and variable $j$, $R_{ij} = 1$ when $x_{ij}$ is missing, and $R_{ij} = 0$ otherwise. We can partition $\bm{x}_j$ as $\bm{x}_j = (\bm{x}_{j}^{obs}, \bm{x}_{j}^{mis})$, where  $\bm{x}_{j}^{obs}$ represents all observed values for variable $j$, i.e., data values in the $\bm{x}_j$ vector corresponding to $R_{ij} = 0$, and $\bm{x}_{j}^{mis}$ represents all missing values for variable $j$, i.e., data values in the $\bm{x}_j$ vector corresponding to $R_{ij} = 1$. We write $\bm{X}^{obs} = (\bm{x}_{1}^{obs},\ldots,\bm{x}_{p}^{obs})$, and $\bm{X}^{mis} = (\bm{x}_{1}^{mis},\ldots,\bm{x}_{p}^{mis})$. Finally, we write $\bm{X} = (\bm{X}^{obs}, \bm{X}^{mis})$.

In this setting, we consider each $\bm{x}_i$ as a group of data containing $p$ categorical responses. This paper assumes that each of these groups has a latent structure, and we wish to find clusters that capture it. In this case, the number of clusters in each group is unknown and we allow clusters to be shared across different groups. Also, we allow our dataset to contain missing values in some variables.

\subsection{The DPMPM model}

There are several existing mixture models that can be applied when we have a categorical dataset with missing values. However, none of them fully takes into account the latent structures that we are dealing with in this article. Among these, the Dirichlet process mixture of products of multinomial distributions (DPMPM, \citet{DPMPM_Kenough}) model is notable due to its flexibility. The DPMPM model assumes that each observation group $i$ in the dataset belongs to a latent class $z_i$ and that variables within each latent class follow independent multinomial distributions. We use the Dirichlet process (DP) as the prior distribution for mixture parameters, where the product of Dirichlet distributions is used as a base probability measure ($G_0$) in this DP. Specifically, let $\theta_{ijd}$ be the probability that $x_{ij}$ takes value $d$, and let $\bm{\theta}_i = \{\theta_{ijd}: j = 1,\ldots,p; d = 1,\ldots,D_j\}$ be the collection of unknown parameters that govern the distribution of $\bm{x}_i$. We can write the DPMPM generative model as  
\begin{align}
G | \alpha_0, G_0 &\sim DP(\alpha_0, G_0)\\
\bm{\theta}_i |G &\stackrel{iid}{\sim} G \quad \text{for all $i$}\\
x_{ij} | \bm{\theta}_i &\stackrel{ind}{\sim} \text{Categorical}(\theta_{ij1},\ldots,\theta_{ijD_j}) \quad \text{ for all $i$ and $j$} \label{DPMPM} 
\end{align}
where $\alpha_0$ is the concentration parameter in the DP prior. Note that $\bm{\theta}_i$ need not be distinct as $G$ from DP is a discrete measure with probability one \citep{blackwell73, seth94}.

An alternative way to illustrate its clustering property is to use the truncated stick-breaking representation of DP \citep{Ishwaran2001} together with a group assignment indicator variable. Let $z_{i} \in \{1,\dots,K\}$ be the cluster assignment of observation $i$, where $K$ is the maximum number of clusters in the truncated stick-breaking process. Let $\phi^{(j)}_{kd} = P(x_{ij} = d|z_{i} = k)$ be the probability that $x_{ij}$ takes value $d$, given that the observation $i$ belongs to a latent class $k$. Finally, we write $\bm{\phi}^{(j)}_{k} = \{\phi^{(j)}_{k1},\ldots,\phi^{(j)}_{kD_j}\}$, $\bm{\phi}_k = \{\bm{\phi}^{(j)}_{k}: j = 1,\ldots,p\}$, and $\bm{\phi} = \{\bm{\phi}_k : k = 1,\ldots,K\}$. We can rewrite the DPMPM generative model as 
\begin{align}
u_k | \alpha_0 &\stackrel{iid}{\sim} \text{Beta}(1,\alpha_0) \text{ for $k$ = 1,\dots,$K$-1}; \quad u_K = 1\\
\pi_{k} &= u_{k}\prod_{h=1}^{k-1}(1-u_{h}); \quad \bm{\phi}_k | G_0 \stackrel{iid}{\sim} G_0 \quad \text{for all $k$}\\
z_i| \pi_1,\ldots,\pi_K &\stackrel{iid}{\sim} \text{Categorical}(\pi_1,\ldots,\pi_K) \quad \text{for all $i$}\\
x_{ij}|z_i, \bm{\phi} &\stackrel{ind}{\sim} \text{Categorical}(\phi^{(j)}_{z_{i}1},\dots,\phi^{(j)}_{z_{i}D_j}) \quad \text{ for all $i$ and $j$}\label{DPMPM-missing} . 
\end{align}
Using this representation, several observations can be associated with the same cluster $k$. With a large enough number of classes $K$, the model is consistent for any joint probability distribution \citep{DPMPM_Kenough}. For more details on the DPMPM model, including further discussions on prior specifications and setting $K$, see \citet{si:reiter:13, manrique:reiter:sm}.

The DPMPM model is flexible as we do not need to know the exact number of clusters in our data beforehand, and we can incorporate missing data imputation into the posterior sampling process using \eqref{DPMPM-missing}. However, it does not exploit the latent structure within each group of observations ${\bm{x}_i}$ that we have, as it assigns the whole observation group to the same mixture component. An alternative would be to have mixture component parameters $G_i$ drawn independently from $DP(\alpha_0, G_0)$ for each observation $i = 1,\dots,n$. However, if $G_0$ is absolutely continuous with respect to the Lebesgue measure, $G_i$ necessarily have no atoms or components in common with one another, and we have no clusters shared between groups. On the other hand, setting the base measure $G_0$ to be discrete would be too restrictive.

\subsection{The GoM model}

Another method commonly utilized in analyzing survey data in social science is the Grade of Membership (GoM, \citet{woodbury:1978, Erosheva:2007}) model. In the GoM modeling, variables in the same observation can be associated with different clusters, and these clusters are shared among groups. Specifically, given the number of clusters $K$, let $z_{ij} \in \{1,\dots,K\}$ be the cluster assignment of $x_{ij}$. Let $\pi_{ik}$, for $k = 1,\ldots,K$, be the probability that a variable of group $i$ will belong to a latent class $k$. Let $\bm{\pi_i} = (\pi_{i1},\dots,\pi_{iK})$. Finally, we redefine $\phi^{(j)}_{kd} = P(x_{ij} = d|z_{ij} = k)$, as variables in the same observation group can now belong to different clusters. We write the generative model as
\begin{align}
\bm{\pi_i} &\stackrel{iid}{\sim} \text{Dirichlet}(\alpha_0\omega_1,\dots,\alpha_0\omega_K) \quad \text{for all $i$}\\
z_{ij} |\bm{\pi}_i &\stackrel{ind}{\sim} \text{Categorical}(\pi_{i1},\ldots,\pi_{iK}) \quad \text{for all $i$ and $j$}\\
x_{ij} | z_{ij}, \bm{\phi} &\stackrel{ind}{\sim} \text{Categorical}(\phi^{(j)}_{z_{ij}1},\dots,\phi^{(j)}_{z_{ij}D_j}) \quad \text{ for all $i$ and $j$} \label{GoM-missing} 
\end{align}
where $\alpha_0$ and $\{\omega_k\}_{k=1}^K$ are model parameters. To fit the model, analysts introduce prior distributions for all model parameters and perform Bayesian inference. For more details on Bayesian inference of the GoM model and its applications, see \citet{Erosheva:2007}. We can incorporate missing data imputation into posterior sampling using \eqref{GoM-missing}.

Even though the GoM model allows latent structures within each group to be exploited and also allows clusters or extreme profiles to be shared across different observation groups, we need to specify the exact number of cluster components $K$ in the model beforehand. In practice, analysts can use prior knowledge of the possible number of profiles or use some criteria such as Akaike's Information Criterion for MCMC samples (AICM) \citep{Raftery2007, Erosheva:2007, gross:2012}. However, fixing the number of components can result in poor estimation of population density. Moreover, especially in the presence of missing values, relying on the ad hoc selection of the number of components can result in using too few components to fully estimate uncertainty in posterior distributions. \citep{si:reiter:13}. This may result in the underestimation of variance in parameter estimates when we perform missing data imputation, which goes against \citet{rubin:1987} recommendations.

\section{The HDPMPM Model} \label{section_HDPMPM}
To overcome the restrictions described in the previous section, this study uses Hierarchical Dirichlet Processes (HDP) introduced by \citet{teh:2006} as the prior for mixture components. Starting with mixture component parameters $G_i$ drawn independently from $DP(\alpha_0, G_0)$ for each observation group $i = 1,\dots,n$, in order to enforce $G_i$ to share atoms across different groups, HDP puts a hyperprior on the global measure $G_0$ as $G_0|\gamma, H \sim DP(\gamma, H)$, where $H$ is a base probability measure of the process and $\gamma$ is a model hyperparameter. As $G_0$ is a realization from a Dirichlet process, it is a discrete measure \citep{blackwell73, seth94}; consequently, $G_i$ for $i = 1,\dots,n$, drawn independently from $DP(\alpha_0, G_0)$, will definitely share atoms with one another. In HDP, we can think of $H$ as the prior distribution for mixture parameters. Then, $G_0$ varies around $H$, with $\gamma$ controlling this variability. Lastly, $G_i$ for each observation group will vary around $G_0$ with the amount of variability controlled by $\alpha_0$.

\subsection{Truncated stick-breaking representation for the HDPMPM model}

Using the truncated stick-breaking representation, the HDPMPM model, which treats each variable as nominal, assumes that $x_{ij}$ belongs to a latent class $z_{ij} \in \{1,\ldots,K\}$, and that variables within each latent class $k = 1,\ldots,K$, follow independent multinomial distributions.

Specifically, let $\pi_{ik} = P(z_{ij} = k)$, for $k = 1,\ldots,K$, be the probability that a variable $j$ from an observation $i$ belongs to a latent class $k$. Let $\phi^{(j)}_{kd} = P(x_{ij} = d|z_{ij} = k)$ be the probability that $x_{ij}$ takes value $d$, given that it belongs to the latent class $k$, as before. Let $\bm{\pi}_i = (\pi_{i1},\ldots,\pi_{iK})$ for each $i$, and $\bm{\pi} =  \{\bm{\pi}_i: i = 1,\ldots,n\}$. We write the generative model as
\begin{align}
z_{ij}| \bm{\pi} &\stackrel{ind}{\sim} \text{Categorical}(\pi_{i1},\ldots,\pi_{iK}) \quad \text{for all $i$ and $j$} \label{HDPMPM:level1}\\
x_{ij}|z_{ij}, \bm{\phi} &\stackrel{ind}{\sim} \text{Categorical}(\phi^{(j)}_{z_{ij}1},\ldots,\phi^{(j)}_{z_{ij}D_j}) \quad \text{for all $i$ and $j$.} \label{HDPMPM:level2} 
\end{align}
Using this model specification, averaging over the $K$ latent classes induces marginal dependence between the variables. This study uses independent flat Dirichlet distributions for each probability vector in $\bm{\phi}$ as the base probability measure $H$. For the prior distribution of the mixture components, we can apply the truncated stick-breaking representation of the Hierarchical Dirichlet Process (HDP) as described by \citet{ren:2008}. Specifically, to obtain $G_0|\gamma, H \sim DP(\gamma, H)$, we have 
\begin{align}
V_k|\gamma &\stackrel{iid}{\sim} \text{Beta}(1, \gamma) \quad \text{ for $k$ = 1,\dots,$K$-1}; \quad V_K = 1\\
\beta_k &= V_k\prod_{h=1}^{k-1}(1-V_h) \quad \text{ for $k$ = 1,\dots,$K$}\\
\bm{\phi}^{(j)}_k &\stackrel{ind}{\sim} \text{Dirichlet}(\bm{1}_{D_j}) \quad \text{ for $j$ = 1,\dots,$p$ and $k$ = 1,\dots,$K$.}
\end{align}
Then, $G_0 = \sum_{k=1}^{K} \beta_{k}\delta_{\bm{\phi}_k}$ where $\delta_{\bm{\phi}_k}$ is a delta function at $\bm{\phi}_k$. We let $\bm{\beta} = (\beta_1,\dots,\beta_K)$ represent the mixture proportions across all groups. To obtain $G_i|\alpha_0, G_0 \stackrel{ind}{\sim} DP(\alpha_0, G_0)$ for each $i = 1,\ldots,n$, we define
\begin{align}
u_{ik}|\alpha_0, \bm{\beta} &\stackrel{ind}{\sim} \text{Beta}(\alpha_0\beta_k, \alpha_0(1-\sum_{h=1}^{k}\beta_h)) \quad \text{for $k$ = 1,\dots,$K$-1}; \quad u_{iK} = 1\\
\pi_{ik} &= u_{ik}\prod_{h=1}^{k-1}(1-u_{ih})  \quad \text{ for $k$ = 1,\dots,$K$}.
\end{align}
Then, $G_i = \sum_{k=1}^{K} \pi_{ik}\delta_{\bm{\phi}_k}$, and $\bm{\pi}_i$ represents the mixture proportions in observation $i$. To complete the model specification, we can define hyperpriors for concentration parameters $\gamma$, and $\alpha_0$ as
\begin{align}
\gamma &\sim Gamma(a, b)\\
\alpha_0 &\sim Gamma(c, d)
\end{align}
where factors $a$, $b$, $c$, and $d$ are positive real constants. To maintain generality and clarity, we use these letters when describing posterior inference. However, in all simulation works, this study follows \citet{dunsonxing} and set $a = b = c = d = 0.25$ to specify vague prior distributions.

\subsection{Gibbs sampling for the HDPMPM model} \label{posterior inference}
In this section, Gibbs sampling scheme for all variables in the HDPMPM model under the truncated stick-breaking representation is presented. The detailed derivation for this scheme is provided in the supplementary material. We note that this method differs from that presented by \citet{teh:2006} as we do not completely integrate out $\bm{\pi}$, which is important for making missing data imputation, and we are using the truncated stick-breaking representation of the Dirichlet process.

The Gibbs sampler starts by first initializing the chain. In the presence of missing values, we need an additional step to set initial values for the missing entries. A popular choice is to set the values by sampling from the marginal distribution of each $\bm{x}_{j}^{obs}$. Alternatively, for each variable, one can also sample from its conditional distribution, given all other variables, where the distribution is constructed using only available cases. Under the HDPMPM model, the model parameters we need to sample from their posterior distributions are $\{z_{ij}\}_{\forall i,j}$, $\{\bm{\phi}^{(j)}_k\}_{\forall j,k}$, $\{\beta_k\}_{\forall k}$, $\{\pi_{ik}\}_{\forall i,k}$, $\gamma$, and $\alpha_0$. We use a dash sign ($-$) in conditional probabilities to represent all data and other parameters. We use $\mathds{1}\{y\}$ to denote an indicator function that equals 1 if $y$ is true and 0 otherwise.

\textbf{Step 1}: For each $x_{ij}$, sample its cluster assignment $z_{ij}$ from a categorical distribution given by
\begin{align}
p(z_{ij}=k|-) &= \frac{\pi_{ik}\phi^{(j)}_{kx_{ij}}}{\sum_{h=1}^{K}\pi_{ih}\phi^{(j)}_{hx_{ij}}} \quad \text{ for $k$ = 1,\dots,$K$}.
\end{align}

\textbf{Step 2}: For $j = 1,\ldots,p$ and $k = 1,\ldots,K$, by using the conjugacy of Dirichlet prior distribution and multinomial likelihood, sample the cluster parameters $\bm{\phi}^{(j)}_k$ from the Dirichlet distribution
\begin{align}
\bm{\phi}^{(j)}_k|- &\stackrel{ind}{\sim} \text{Dirichlet}(1+n^{(j)}_{k1},\dots,1+n^{(j)}_{kD_j})
\end{align}
where $n^{(j)}_{kd}=\sum_{i=1}^{n} \mathds{1}\{z_{ij}=k \wedge x_{ij}=d\}$ is the number of responses of variable $j$ in the dataset that belong to cluster $k$ and take on value $d$ in the current iteration. 

\textbf{Step 3}: Update the global mixture proportions $(\beta_1,\ldots,\beta_K)$ by using an auxiliary variable approach where $t_i$ for $i = 1,\ldots,n$ and $s_{ik}$ for $i = 1,\ldots,n$ and $k = 1,\ldots,K$ are introduced as auxiliary variables. Let $m_{ik} = \sum_{j=1}^{p} \mathds{1}\{z_{ij}=k\}$ be the number of members in observation group $i$ that belong to cluster $k$. We have 
\begin{align}
t_i | - &\stackrel{ind}{\sim} \text{Beta}(\alpha_0,\sum_{k=1}^{K}m_{ik}) \quad \text{ for $i$ = 1,\dots,$n$}\\
s_{ik}^{(h)} | - &\stackrel{ind}{\sim} \text{Bernoulli}(\frac{\alpha_0\beta_k}{\alpha_0\beta_k +h -1}) \quad \text{for $h$ = 1,\dots,$m_{ik}$; $k$ = 1,\dots,$K$; $i$ = 1,\dots,$n$}\\
s_{ik} &= \sum_{h=1}^{m_{ik}} s_{ik}^{(h)} \quad \text{ for $i$ = 1,\dots,$n$; $k$ = 1,\dots,$K$}\\
V_k | - &\stackrel{ind}{\sim} \text{Beta}(1+\sum_{i=1}^{n}s_{ik}, \gamma + \sum_{i=1}^{n}\sum_{h=k+1}^{K}s_{ih}) \quad \text{ for $k$ = 1,\dots,$K$-1}; \quad V_K = 1\\
\beta_k &= V_k\prod_{h=1}^{k-1}(1-V_h) \quad \text{ for $k$ = 1,\dots,$K$}.
\end{align}

\textbf{Step 4}: For each $i = 1,\ldots,n$, update the group mixture proportions $\bm{\pi}_i$. As, in HDP, the prior distribution of this variable is $\bm{\pi}_i|\alpha_0, \bm{\beta} \stackrel{iid}{\sim} DP(\alpha_0, \bm{\beta})$ \citep{teh:2006}, we have 
\begin{align}
u_{ik}|- &\stackrel{ind}{\sim} \text{Beta}(\alpha_0\beta_k+m_{ik}, \alpha_0(1-\sum_{h=1}^{k}\beta_h)+\sum_{h=k+1}^{K}m_{ih}) \quad \text{for $k$ = 1,\dots,$K$-1}; \quad u_{iK} = 1\\
\pi_{ik} &= u_{ik}\prod_{h=1}^{k-1}(1-u_{ih})  \quad \text{ for $k$ = 1,\dots,$K$}.
\end{align}

\textbf{Step 5}: Sample concentration parameters $\gamma$ and $\alpha_0$ using the conditional distributions
\begin{align}
\gamma|- &\sim \text{Gamma}(a+K-1, b-\sum_{k=1}^{K-1}\text{ln}(1-V_k))\\
\alpha_0|- &\sim \text{Gamma}(c + \sum_{i=1}^{n}\sum_{k=1}^{K}s_{ik}, d - \sum_{i=1}^{n}\text{ln}(t_i).
\end{align}

\textbf{Step 6}: Handle missing values directly within the Gibbs sampler. At any iteration $t$, we sample each $x_{ij}$ with $R_{ij} = 1$ using \eqref{HDPMPM:level2} conditional on the current draw of the parameters and its latent class $z_{ij}$ as follows
\begin{align}
x_{ij}|- &\stackrel{ind}{\sim} \text{Categorical}(\phi^{(j)}_{z_{ij}1},\ldots,\phi^{(j)}_{z_{ij}D_j}) \quad \text{for all $i$ and $j$ where $R_{ij}=1$}.
\end{align}

To set the maximum number of clusters in the truncated stick-breaking process, this study follows \citet{si:reiter:13}. Starting with a moderate value of $K$ during initial runs, whenever the number of occupied clusters reaches this upper limit in any iteration, we gradually increase $K$ and rerun the MCMC until this is no longer the case.

\section{Application to the 2016 ANES Survey Data} \label{section_ANES_appl}

\subsection{Framework and the ANES data}

 An application of the HDPMPM modeling is demonstrated using ANES 2016 time series study data from \citet{ANES2016}. This survey data is part of the series of election studies conducted since 1948 to provide data for the analysis of public opinion and voting behavior in the U.S. presidential elections. The 2016 survey included responses from pre-election interviews and post-election interviews of the same respondents. The total pre-election sample size was 4,271. These samples were drawn independently from U.S. citizens aged 18 or older using address-based sampling. The data include over 1,800 variables covering different aspects, including voting behavior, candidate and party evaluations, government evaluations, political issues, etc.  The data can be downloaded from the Inter-university Consortium for Political and Social Research (ICPSR) website \textit{(https://www.icpsr.umich.edu/web/ICPSR/studies/36824/summary}).

For this study, we combine pre-election responses and post-election responses of the same individuals and treat this as our unit of analysis. All observations that did not respond to the post-election survey are removed. This study focuses only on a subset of categorical variables from this dataset, mostly pertaining to political issues and opinions related to the three core beliefs in American political ideology: support for equality of opportunity, economic individualism, and the free enterprise system \citep{feldman:1988}. This results in a dataset containing 3,409 records and 23 categorical variables, which we treat as our population data. To avoid overparameterization with a small sample size and to further simplify the analysis, this study follows the data preprocessing steps in \citet{gross:2012} and merge similar levels of variables to have three levels per variable. Table \ref{data_dictionary} reports the data dictionary. The variables are described in more detail in the supplementary material.

\begin{singlespace}
\begin{footnotesize}
\begin{longtable}{llp{5cm}p{8.5cm}}
\caption{Data dictionary of our population dataset.} \label{data_dictionary} \\
  \toprule 
 No. & Variables & Description & Levels \\ 
  \midrule
  \endfirsthead
\toprule 
 No. & Variables & Description & Levels \\ 
  \midrule
  \endhead
  \bottomrule
  \endlastfoot
  1& V162123 & Agree/disagree: Better if rest of world more like America & 1=Agree, 2=Neither agree nor disagree, 3=Disagree\\
  2& V162134 & How much opportunity is there in America today to
get ahead? & 1=A great deal/A lot, 2=A moderate amount/A little, 3=None\\ 
  3& V162140 & Favor or oppose tax on millionaires & 1=Favor, 2=Oppose, 3=Neither favor nor oppose\\ 
  4& V162145 & Health Care Law effect on cost of respondent's health care & 1=Increased, 2=Decreased, 3=Had no effect\\ 
  5& V162148 & Favor or oppose government reducing income inequality
 & 1=Favor, 2=Oppose, 3=Neither favor nor oppose\\ 
  6& V162158 &  How likely immigration will take away jobs? & 1=Extremely likely, 2=Very likely/Somewhat likely, \newline 3=Not at all likely\\ 
  7& V162170 &  Agree/disagree: Country needs strong leader to take us back to true path & 1=Agree, 2=Neither agree nor disagree, 3=Disagree\\ 
  8& V162176 & Favor or oppose free trade agreements with other countries & 1=Favor, 2=Oppose, 3=Neither favor nor oppose\\ 
  9& V162179 & Should marijuana be legal? & 1=Favor, 2=Oppose, 3=Neither favor nor oppose\\ 
  10& V162180 & Should the government do more or less to regulate banks? & 1=More, 2=Less, 3=The same\\ 
  11& V162192 & Should the minimum wage be raised? & 1=Raised, 2=Kept the same, 3=Lowered/Eliminated\\ 
  12& V162193 & Increase or decrease government spending to help people pay for health care & 1=Increase, 2=Decrease, 3=No change\\
  13& V162207 & Agree/disagree: world is changing and we should adjust & 1=Agree, 2=Neither agree nor disagree, 3=Disagree\\
  14& V162208 & Agree/disagree: newer lifestyles breaking down society & 1=Agree, 2=Neither agree nor disagree, 3=Disagree\\
  15& V162209 & Agree/disagree: We should be more tolerant of other moral standards & 1=Agree, 2=Neither agree nor disagree, 3=Disagree\\
  16& V162212 & Agree/disagree: past slavery and discrimination made it difficult for
blacks to work their way out of the lower class & 1=Agree, 2=Neither agree nor disagree, 3=Disagree\\
  17& V162214 & Agree/disagree: if blacks would only try harder, they
could be just as well off as whites & 1=Agree, 2=Neither agree nor disagree, 3=Disagree\\
  18& V162231 & Should the news media pay more attention to discrimination against women? & 1=More attention, 2=Less attention, \newline 3=Same amount of attention\\
  19& V162246 & Agree/disagree: If people were
treated more equally in this country, we would have many fewer problems & 1=Agree, 2=Neither agree nor disagree, 3=Disagree\\
  20& V162260 & Agree/disagree: Most politicians do not care about the people & 1=Agree, 2=Neither agree nor disagree, 3=Disagree\\
  21& V162269 & Agree/disagree: America’s culture is generally harmed by immigrants & 1=Agree, 2=Neither agree nor disagree, 3=Disagree\\
  22& V162271 & To be truly American, it is important to have been born in U.S. & 1=Very important,\newline 2=Fairly important/Not very important,\newline 3=Not important at all\\
  23& V162290 & Satisfied with way democracy works in the U.S. & 1=Very satisfied, 2=Fairly satisfied/Not very satisfied,\newline 3=Not at all satisfied\\
   \hline 
\end{longtable}
\end{footnotesize}
\end{singlespace}

We first use the HDPMPM in modeling our dataset when it is fully observed. To implement the HDPMPM model, the author uses his own \textsf{R} code to perform Gibbs sampling to generate posterior draws of the parameters as described in section \ref{posterior inference}. The maximum number of latent classes is set to $K=30$, based on initial runs on the dataset. As the HDPMPM model is invariant to permutations of the cluster labels, to alleviate this label switching, clusters are relabeled in each iteration of MCMC according to the decreasing sequence of posterior samples of their proportions in $\bm{\beta}$. The Gibbs sampler was run for 30,000 iterations, with the first 15,000 iterations designated as burn-in. The autocorrelation in posterior samples was reduced by thinning the posterior samples with a factor of 5. The convergence of the sampler was assessed by examining trace plots of combinations of the model parameters, such as random samples of marginal probabilities, that are insensitive to label switching.

\subsection{HDPMPM inference on fully observed ANES dataset}
This section summarizes the results of modeling our dataset from the 2016 ANES data with the HDPMPM model using graphical displays and tables. The most important findings for the fully observed scenario, including insights of extreme profiles in the population data, is presented. We also inspect and discuss different issues commonly found in applications of Bayesian mixture models.  

\begin{figure}[t!]
\centerline{\includegraphics[scale=.7]{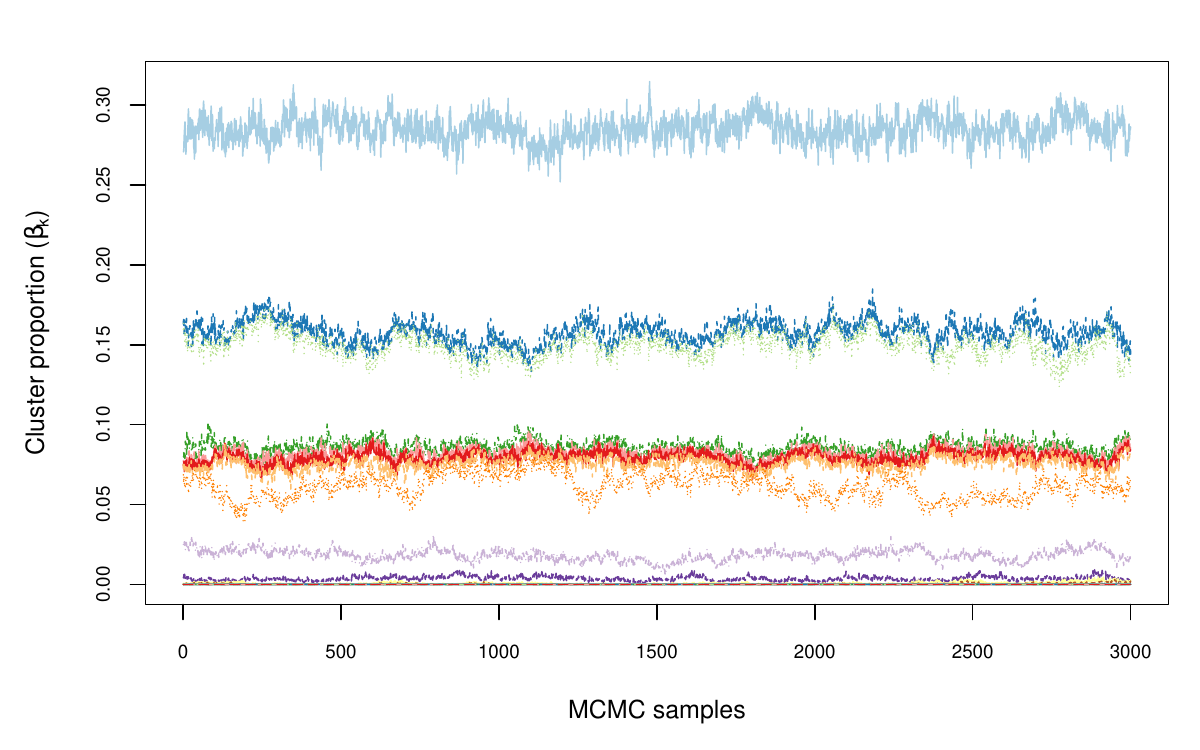}}
\caption{Trace plot of posterior samples of population-level cluster proportions $\beta_k$.}
\label{post_beta}
\end{figure}

Figure \ref{post_beta} displays the trace plot of the posterior samples of population-level proportions of different profiles $\beta_k$ for $k = 1,\ldots,30$. Visually, we are able to obtain convergence of MCMC with good mixing behavior in the posterior samples. There are three clusters with expected population proportions over 0.1. The response profiles $\{\bm{\phi}_k^{(j)}\}_{\forall j}$ for these three clusters from posterior inference do not show significant label switching behavior. Hence, we choose to analyze and describe them further.

We note that the theoretical support for using the Dirichlet process mixture in clustering and inferring the number of mixture components is limited. The Dirichlet process can generate a number of small clusters which may not reflect the underlying data generating process \citep{Green2001,Miller2014}. Moreover, the model  is inconsistent in the posterior number of clusters under some mild assumptions \citep{Miller2013,Miller2014}. Therefore, this work focuses the main analysis only on large mixture components.

To simplify terminology and to connect our analysis with common notions of political ideology, we will call the biggest cluster in this simulation ($k=1$) the \textit{liberal} group and the second largest cluster ($k=2$) the \textit{conservative} group. We note that these political identifiers are not absolute. Moreover, political scientists may argue that these political profiles occur on a continuous spectrum rather than discrete categories \citep{maddox:1984, feldman:2009}. As accurately characterizing political prototypes is not the main goal of this study, we will use the identifiers, \textit{liberal} and \textit{conservative}, in only a relative sense.

By analyzing posterior expectations of person-level cluster proportions $\bm{\pi}_i$, about 34\% of individuals in our population are primarily associated with one cluster, where the cluster's contribution exceeds 10\% of their total profile. It is noted that a significant number of people are associated with more than one dominant profile. Specifically, 30\% and 24\% of our population are associated mainly with two and three extreme profiles, respectively. Looking at the most dominant profile in each person, 32\% of our population has the \textit{liberal} profile as their dominant profile, while 17\% have the \textit{conservative} profile as their most probable political profile.

Table \ref{extreme-profiles} reports the posterior expectations of the \textit{liberal} and the \textit{conservative} extreme profiles. The posterior standard deviations of the \textit{liberal} profile range from 0.002 to 0.021, while for the \textit{conservative} profile, they range from 0.006 to 0.037.

We can justify the names given to the two profiles by analyzing these posterior expectations. For example, regarding the government policy, our \textit{liberal} group favors (with probability 0.88) an increase in government spending to help people pay for health insurance when they cannot afford it (V162193). In contrast, the \textit{conservative} group wants to decrease the government spending on it (with probability 0.80). Consistent with the contemporary definition of conservatism, our prototypical \textit{conservatives} tend to disagree (with probability 0.61) with the statement, ‘We should be more tolerant of people who choose to live according to their own moral standards, even if they are very different from our own’ (V162209), and admit (with probability 0.98) that newer lifestyles are contributing to the breakdown of our society (V162208) while the prototypical \textit{liberals} will favor the opposite direction. Lastly, from the view of locus of responsibility, the \textit{conservatives} would put more emphasis on individual responsibility for their success. For example, our \textit{conservatives} would oppose (with probability 0.95) the government's role in trying to reduce the difference in incomes between the rich and the poor (V162148), while the \textit{liberals} would show support (with probability 0.82) for such intervention. Our \textit{conservative} profile would agree (with probability 0.84) that black people can be as well off as white people if they only try hard enough (V162214), while the \textit{liberals} would believe (with probability 0.97) that it is not a matter of whether black people are trying hard enough or not. However, the author admits that this racial argument is subject to biases of different groups of people toward people of other races. 

For the third dominant cluster, it is a mix between the \textit{liberal} profile and the \textit{conservative} profile, with less extreme dominant responses. In some survey questions, the patterns are similar to the \textit{conservative} group but with more uniform answers, while others may look like mild \textit{liberal} patterns. This may represent opinions from a prototypical ideology in the middle of the political spectrum.

\begin{table}[t!]
\centering
\footnotesize
\caption{Posterior expectations of the \textit{liberal} and the \textit{conservative} extreme profiles from fully observed ANES dataset.} 
\begin{tabular}{lrrr|rrr}
  \toprule
\multicolumn{1}{c}{}  &   \multicolumn{3}{c}{Liberal} & \multicolumn{3}{c}{Conservative}\\
\multicolumn{1}{c}{}  &   \multicolumn{3}{c}{proportion 28.49\%} & \multicolumn{3}{c}{proportion 15.77\%}\\
\hline
& level 1 & level 2 & level 3 & level 1 & level 2 & level 3 \\ 
\hline
V162123: Better if rest of world more like U.S. & 0.05 & 0.20 & 0.76 & 0.53 & 0.38 & 0.09 \\ 
  V162134: How much opportunity is there in U.S. & 0.15 & 0.83 & 0.02 & 0.49 & 0.49 & 0.02 \\ 
  V162140: Favor/oppose tax on millionaires & 0.99 & 0.00 & 0.01 & 0.14 & 0.52 & 0.34 \\ 
  V162145: Health Care Law effect on cost & 0.27 & 0.12 & 0.61 & 0.79 & 0.02 & 0.19 \\ 
  V162148: Favor/oppose reduce income inequality & 0.82 & 0.03 & 0.15 & 0.01 & 0.95 & 0.04 \\ 
  V162158: How likely immigrants take away jobs & 0.00 & 0.37 & 0.62 & 0.34 & 0.63 & 0.02 \\ 
  V162170: Country needs strong leader & 0.03 & 0.13 & 0.84 & 0.91 & 0.07 & 0.02 \\ 
  V162176: Favor/oppose free trade agreements & 0.65 & 0.09 & 0.26 & 0.35 & 0.37 & 0.28 \\ 
  V162179: Should marijuana be legal & 0.80 & 0.04 & 0.16 & 0.17 & 0.64 & 0.20 \\ 
  V162180: Should regulate banks more or less  & 0.83 & 0.01 & 0.16 & 0.14 & 0.48 & 0.37 \\ 
  V162192: Should the minimum wage be raised & 0.97 & 0.03 & 0.00 & 0.07 & 0.71 & 0.23 \\ 
  V162193: Government pay people's health care & 0.88 & 0.00 & 0.12 & 0.01 & 0.80 & 0.20 \\ 
  V162207: World changing and we should adjust & 0.75 & 0.07 & 0.18 & 0.06 & 0.03 & 0.91 \\ 
  V162208: Newer lifestyles break down society & 0.06 & 0.07 & 0.87 & 0.98 & 0.01 & 0.01 \\ 
  V162209: More tolerant of other moral standards & 0.95 & 0.03 & 0.01 & 0.19 & 0.19 & 0.61 \\ 
  V162212: Past slavery made it difficult for blacks & 0.96 & 0.02 & 0.02 & 0.10 & 0.01 & 0.89 \\ 
  V162214: Blacks must try harder to get ahead & 0.00 & 0.03 & 0.97 & 0.84 & 0.13 & 0.04 \\ 
  V162231: More media attention to discrimination & 0.85 & 0.00 & 0.15 & 0.03 & 0.63 & 0.35 \\ 
  V162246: Fewer problems if treated people fairly & 0.96 & 0.03 & 0.01 & 0.20 & 0.27 & 0.53 \\ 
  V162260: Most politicians not care about people & 0.47 & 0.11 & 0.41 & 0.65 & 0.12 & 0.24 \\ 
  V162269: U.S.’s culture is harmed by immigrants & 0.00 & 0.00 & 1.00 & 0.40 & 0.31 & 0.28 \\ 
  V162271: Born in U.S. to be truly American  & 0.01 & 0.49 & 0.50 & 0.28 & 0.62 & 0.10 \\ 
  V162290: Satisfied with democracy in U.S. & 0.02 & 0.89 & 0.09 & 0.19 & 0.78 & 0.03 \\ 
  \bottomrule
\end{tabular}
\label{extreme-profiles}
\end{table}


With posterior samples from the HDPMPM model, there are several political science tools we can use to identify items that have high or low valence in separating these extreme profiles. Two such tools, discussed in \citet{gross:2012}, are the cohesion ratio and the disagreement score. \citet{gross:2012} defined the cohesion ratio of the extreme profile $k$ with respect to variable $j$ as
\begin{align}
CR_{jk} = \frac{\underset{d=1,\ldots,D_j}{\max}\phi_{kd}^{(j)}}{\underset{d=1,\ldots,D_j}{\min}\phi_{kd}^{(j)}}
\end{align}
which is unbounded above. To restrict the upper bound and make comparison simpler, we redefine the cohesion ratio ($CR_{jk}$) as
\begin{align}
CR_{jk} = \frac{\underset{d=1,\ldots,D_j}{\max}\phi_{kd}^{(j)} - \underset{d=1,\ldots,D_j}{\min}\phi_{kd}^{(j)}}{\underset{d=1,\ldots,D_j}{\max}\phi_{kd}^{(j)}}
\end{align}
which takes values from 0 to 1. This score gives a rough measure of how predictable the response for variable $j$ is given that individuals are drawing conclusions from profile $k$ \citep{gross:2012}. Note that the two definitions give the same information about the relative proportions of the most dominant and the least dominant responses.

Another useful tool is the disagreement score ($DR$) between profiles $k_1$ and $k_2$ for an item $j$. This paper uses the same definition given in \citet{gross:2012}: 
\begin{align}
DR_j = \mathds{1} \{\underset{d=1,\ldots,D_j}{\text{arg\,max}}\, \phi_{k_1d}^{(j)} \neq \underset{d=1,\ldots,D_j}{\text{arg\,max}}\, \phi_{k_2d}^{(j)}\}.
\end{align}
Hence, the posterior expectation of $DR_j$ gives us a way of measuring how strongly variable $j$ distinguishes the two profiles in terms of their most dominant responses to that variable.

Table \ref{cohesion-disagreement} presents the posterior expectations of the cohesion ratio and the disagreement score between our \textit{liberal} and \textit{conservative} profiles. Unlike the results obtained from working with survey responses on political core values in \citet{gross:2012}, our \textit{liberal} and \textit{conservative} profiles created from the subset of the ANES survey mostly disagree with each other, with few exceptions. For example, both groups share a similar modal response of agreeing (with probability 0.47 for the \textit{liberal} profile and 0.65 for the \textit{conservative} profile) that most politicians do not care about the people (V162260). However, the pure \textit{conservatives} show higher cohesion and hence would be more adherent to this response. Even though the two profiles do not entirely disagree on the idea that it is important to have been born in the U.S. to be truly American (V162271), the \textit{liberal} profile is slightly more cohesive in their opinion. This is obvious considering that there is only a 0.01 probability of \textit{liberals} claiming that it is actually important to have been born in the U.S., whereas the responses from the \textit{conservative} profile are more split into all three levels. Lastly, both profiles tend to show the same modal response when asked whether they are satisfied with the way democracy works in the U.S. (V162290), with the same level of cohesion ratio. 

\begin{table}[t!]
\centering
\footnotesize
\caption{Posterior expectations of the cohesion ratio and the disagreement score from fully observed ANES dataset.} 
\begin{tabular}{lrr|r}
  \toprule
\multicolumn{1}{c}{}  &   \multicolumn{2}{c}{Cohesion Ratio} & \multicolumn{1}{c}{Disagreement}\\
\multicolumn{1}{c}{}  &   \multicolumn{1}{c}{Liberal} & \multicolumn{1}{c}{Conservative} & \multicolumn{1}{c}{}\\
\hline
V162123: Better if rest of world more like U.S. & 0.94 & 0.83 & 1.00 \\ 
  V162134: How much opportunity is there in U.S. & 0.98 & 0.95 & 0.51 \\ 
  V162140: Favor/oppose tax on millionaires & 1.00 & 0.72 & 1.00 \\ 
  V162145: Health Care Law effect on cost & 0.81 & 0.97 & 1.00 \\ 
  V162148: Favor/oppose reduce income inequality & 0.96 & 0.99 & 1.00 \\ 
  V162158: How likely immigrants take away jobs & 1.00 & 0.96 & 1.00 \\ 
  V162170: Country needs strong leader & 0.97 & 0.98 & 1.00 \\ 
  V162176: Favor/oppose free trade agreements & 0.86 & 0.27 & 0.63 \\ 
  V162179: Should marijuana be legal & 0.95 & 0.74 & 1.00 \\ 
  V162180: Should regulate banks more or less & 0.99 & 0.70 & 1.00 \\ 
  V162192: Should the minimum wage be raised & 1.00 & 0.90 & 1.00 \\ 
  V162193: Government pay people's health care & 1.00 & 0.99 & 1.00 \\ 
  V162207: World changing and we should adjust & 0.90 & 0.97 & 1.00 \\ 
  V162208: Newer lifestyles break down society & 0.94 & 0.99 & 1.00 \\ 
  V162209: More tolerant of other moral standards & 0.99 & 0.71 & 1.00 \\ 
  V162212: Past slavery made it difficult for blacks & 0.99 & 0.98 & 1.00 \\ 
  V162214: Blacks must try harder to get ahead & 1.00 & 0.96 & 1.00 \\ 
  V162231: More media attention to discrimination & 1.00 & 0.96 & 1.00 \\ 
  V162246: Fewer problems if treated people fairly & 0.99 & 0.62 & 1.00 \\ 
  V162260: Most politicians not care about people & 0.76 & 0.82 & 0.07 \\ 
  V162269: U.S.’s culture is harmed by immigrants & 1.00 & 0.32 & 0.97 \\ 
  V162271: Born in U.S. to be truly American & 0.98 & 0.84 & 0.55 \\ 
  V162290: Satisfied with democracy in U.S. & 0.97 & 0.97 & 0.00 \\ 
  \bottomrule
\end{tabular}
\label{cohesion-disagreement}
\end{table}

These insightful characteristics of dominant profiles are obtained consistently even with the randomness in the clustering property of the Dirichlet process mixture. The entire MCMC sampling process was rerun many times using different random seeds to initialize the chain. The number of moderate and large clusters varies from three up to five dominant profiles, among which profiles similar to our \textit{conservative} and \textit{liberal} profiles previously shown always appear. Upon closer inspection, the other profiles are not as extreme as the \textit{conservative} and the \textit{liberal} ones in terms of their mutual disagreement and the cohesion toward their responses. Therefore, we can recognize these as profiles from which someone in the middle of the political spectrum may draw conclusions about different political issues.

In some very rare cases, due to the statistical nature of the Dirichlet process in clustering, we may observe that an original extreme profile breaks down into smaller sub-profiles. For example, Table \ref{group-splitting-behavior} shows this splitting behavior in the \textit{liberal} group observed in one of the realizations of the HDPMPM modeling. Clearly, the two sub-profiles are considerably smaller in proportion compared to what we obtained previously for the \textit{liberal} profile. Moreover, the two sub-profiles share the same modal response pattern in most of the items and vary only slightly from the \textit{liberal} profile we obtained earlier.

\begin{table}[t!]
\centering
\scriptsize
\caption{Splitting behavior of the \textit{liberal} profile from fully observed ANES dataset. To make comparison easier, the posterior expectations of the original \textit{liberal} profile is reported on the left side of this table.} 
\begin{tabular}{lrrr|rrr|rrr}
  \toprule
\multicolumn{1}{c}{}  &  \multicolumn{3}{c}{Liberal profile} &  \multicolumn{6}{c}{Liberal sub-profiles}\\
\multicolumn{1}{c}{}  &  \multicolumn{3}{c}{} & \multicolumn{3}{c}{sub-profile 1} & \multicolumn{3}{c}{sub-profile 2} \\
\multicolumn{1}{c}{}  &  \multicolumn{3}{c}{proportion 28.49\%} & \multicolumn{3}{c}{proportion 11.57\%} & \multicolumn{3}{c}{proportion 10.90\%} \\
\hline
& level 1 & level 2 & level 3 & level 1 & level 2 & level 3  & level 1 & level 2 & level 3\\
\hline
V162123 & 0.05 & 0.20 & 0.76 & 0.11 & 0.21 & 0.69 & 0.01 & 0.17 & 0.82 \\ 
  V162134 & 0.15 & 0.83 & 0.02 & 0.30 & 0.69 & 0.00 & 0.04 & 0.92 & 0.04 \\ 
  V162140 & 0.99 & 0.00 & 0.01 & 0.97 & 0.01 & 0.02 & 0.98 & 0.01 & 0.01 \\ 
  V162145 & 0.27 & 0.12 & 0.61 & 0.24 & 0.10 & 0.66 & 0.30 & 0.13 & 0.57 \\ 
  V162148 & 0.82 & 0.03 & 0.15 & 0.64 & 0.09 & 0.27 & 0.93 & 0.01 & 0.05 \\ 
  V162158 & 0.00 & 0.37 & 0.62 & 0.00 & 0.39 & 0.61 & 0.00 & 0.32 & 0.68 \\ 
  V162170 & 0.03 & 0.13 & 0.84 & 0.02 & 0.11 & 0.87 & 0.04 & 0.12 & 0.84 \\ 
  V162176 & 0.65 & 0.09 & 0.26 & 0.80 & 0.01 & 0.19 & 0.50 & 0.18 & 0.31 \\ 
  V162179 & 0.80 & 0.04 & 0.16 & 0.69 & 0.05 & 0.26 & 0.92 & 0.03 & 0.05 \\ 
  V162180 & 0.83 & 0.01 & 0.16 & 0.67 & 0.01 & 0.31 & 0.94 & 0.01 & 0.05 \\ 
  V162192 & 0.97 & 0.03 & 0.00 & 0.89 & 0.11 & 0.00 & 0.99 & 0.01 & 0.00 \\ 
  V162193 & 0.88 & 0.00 & 0.12 & 0.78 & 0.01 & 0.21 & 0.94 & 0.01 & 0.05 \\ 
  V162207 & 0.75 & 0.07 & 0.18 & 0.70 & 0.07 & 0.23 & 0.80 & 0.08 & 0.12 \\ 
  V162208 & 0.06 & 0.07 & 0.87 & 0.03 & 0.05 & 0.93 & 0.06 & 0.05 & 0.89 \\ 
  V162209 & 0.95 & 0.03 & 0.01 & 0.96 & 0.02 & 0.01 & 0.93 & 0.04 & 0.02 \\ 
  V162212 & 0.96 & 0.02 & 0.02 & 0.90 & 0.01 & 0.09 & 0.97 & 0.02 & 0.01 \\ 
  V162214 & 0.00 & 0.03 & 0.97 & 0.02 & 0.03 & 0.95 & 0.01 & 0.03 & 0.96 \\ 
  V162231 & 0.85 & 0.00 & 0.15 & 0.71 & 0.01 & 0.28 & 0.93 & 0.01 & 0.06 \\ 
  V162246 & 0.96 & 0.03 & 0.01 & 0.89 & 0.06 & 0.05 & 0.98 & 0.01 & 0.01 \\ 
  V162260 & 0.47 & 0.11 & 0.41 & 0.28 & 0.11 & 0.61 & 0.65 & 0.10 & 0.25 \\ 
  V162269 & 0.00 & 0.00 & 1.00 & 0.00 & 0.00 & 0.99 & 0.01 & 0.01 & 0.99 \\ 
  V162271 & 0.01 & 0.49 & 0.50 & 0.01 & 0.44 & 0.55 & 0.01 & 0.48 & 0.50 \\ 
  V162290 & 0.02 & 0.89 & 0.09 & 0.06 & 0.93 & 0.01 & 0.01 & 0.82 & 0.17 \\ 
  \bottomrule
\end{tabular}
\label{group-splitting-behavior}
\end{table}

In this case, by using domain knowledge and the posterior samples of their respective proportions $\beta_k$, we can merge the sub-profiles and recover the larger one if needed. Table \ref{merge-split-groups} numerically reports the extreme profile obtained by merging these two \textit{liberal} sub-profiles. We have a nearly perfect match between the \textit{liberal} profile shown previously and the merged profile obtained by combining the two sub-profiles in Table \ref{group-splitting-behavior}, even though we started the posterior sampling with a totally different initialization of the MCMC. We note that the decision to merge smaller profiles into larger ones is subjective and can largely depend on the domain of study. Analysts can also use algorithms such as the one proposed in \citet{guha2019posterior}, which relies on the posterior contraction of the mixing measure, to merge similar clusters together. However, the discussion on the criteria for making this decision is beyond the scope of our study. For more details on this topic, see \citet{Leroux1992, Miller2014,guha2019posterior}.

\begin{table}[t!]
\centering
\footnotesize
\caption{Comparison between the \textit{liberal} profile and the one obtained from merging similar \textit{liberal} sub-profiles.} 
\begin{tabular}{lrrr|rrr}
  \toprule
\multicolumn{1}{c}{} & \multicolumn{3}{c}{Liberal profile} &   \multicolumn{3}{c}{Merged profile} \\
\multicolumn{1}{c}{} & \multicolumn{3}{c}{proportion 28.49\%} &   
\multicolumn{3}{c}{proportion 22.47\%} \\
\hline
& level 1 & level 2 & level 3 & level 1 & level 2 & level 3 \\ 
\hline
V162123 & 0.05 & 0.20 & 0.76 & 0.06 & 0.19 & 0.75 \\ 
  V162134 & 0.15 & 0.83 & 0.02 & 0.17 & 0.80 & 0.02 \\ 
  V162140 & 0.99 & 0.00 & 0.01 & 0.98 & 0.01 & 0.02 \\ 
  V162145 & 0.27 & 0.12 & 0.61 & 0.27 & 0.11 & 0.62 \\ 
  V162148 & 0.82 & 0.03 & 0.15 & 0.78 & 0.05 & 0.16 \\ 
  V162158 & 0.00 & 0.37 & 0.62 & 0.00 & 0.36 & 0.64 \\ 
  V162170 & 0.03 & 0.13 & 0.84 & 0.03 & 0.12 & 0.85 \\ 
  V162176 & 0.65 & 0.09 & 0.26 & 0.66 & 0.10 & 0.25 \\ 
  V162179 & 0.80 & 0.04 & 0.16 & 0.80 & 0.04 & 0.16 \\ 
  V162180 & 0.83 & 0.01 & 0.16 & 0.80 & 0.01 & 0.19 \\ 
  V162192 & 0.97 & 0.03 & 0.00 & 0.94 & 0.06 & 0.00 \\ 
  V162193 & 0.88 & 0.00 & 0.12 & 0.86 & 0.01 & 0.13 \\ 
  V162207 & 0.75 & 0.07 & 0.18 & 0.75 & 0.07 & 0.18 \\ 
  V162208 & 0.06 & 0.07 & 0.87 & 0.05 & 0.05 & 0.91 \\ 
  V162209 & 0.95 & 0.03 & 0.01 & 0.95 & 0.03 & 0.02 \\ 
  V162212 & 0.96 & 0.02 & 0.02 & 0.93 & 0.02 & 0.05 \\ 
  V162214 & 0.00 & 0.03 & 0.97 & 0.01 & 0.03 & 0.95 \\ 
  V162231 & 0.85 & 0.00 & 0.15 & 0.82 & 0.01 & 0.17 \\ 
  V162246 & 0.96 & 0.03 & 0.01 & 0.93 & 0.04 & 0.03 \\ 
  V162260 & 0.47 & 0.11 & 0.41 & 0.46 & 0.11 & 0.43 \\ 
  V162269 & 0.00 & 0.00 & 1.00 & 0.00 & 0.00 & 0.99 \\ 
  V162271 & 0.01 & 0.49 & 0.50 & 0.01 & 0.46 & 0.53 \\ 
  V162290 & 0.02 & 0.89 & 0.09 & 0.04 & 0.88 & 0.09 \\ 
  \bottomrule
\end{tabular}
\label{merge-split-groups}
\end{table}

\section{HDPMPM Inference with Missing Data} \label{section_missing_data}

This section illustrates the use of the HDPMPM in modeling data with missing values using our population dataset from the 2016 ANES data. We explore two different scenarios in our dataset: (i) values missing at random (MAR, \citet{littlerubin}), and (ii) values missing completely at random (MCAR, \citet{littlerubin}). To create missing data under both MAR and MCAR, firstly, eleven of the variables was set to be fully observed. This mimics real applications where some variables are often fully observed. Next, 30\% of the values of the remaining twelve variables was set to be missing according to either MCAR or MAR mechanisms. These scenarios tend to result in less than 5\% complete cases, making complete case analysis untenable.

To implement the HDPMPM model with missing values, this study performs Gibbs sampling to generate posterior draws of the parameters and handles missing values directly within the sampler as described in section \ref{posterior inference}. The following sections report the analysis of their extreme profiles and compare the results with those obtained from the fully observed dataset.

\subsection{MAR scenario}
We first consider the MAR scenario as it is often the most likely missing data mechanism that analysts need to deal with in practice. To incorporate MAR into our population dataset, missing values is set independently for six variables (V162123, V162148, V162170, V162192, V162209, and V162246) with a 30\% missingness rate. Then, missingness indicators are generated for six other variables (V162140, V162158, V162179, V162207, V162214, and V162269) using logistic regression models, conditional on the remaining eleven fully observed variables. For simplicity, only the main effects are included in these logistic regression models. The coefficients of the main effects are adjusted to result in approximately $30\%$ missing data per variable. The supplementary material presents these logistic regression models.

By analyzing the posterior samples of the population-level proportions of different extreme profiles $\beta_k$, there are four clusters with expected population proportions over 0.1. The number of large extreme profiles may increase due to the additional variability in the model caused by missing data. However, the number is close to what we have for the HDPMPM modeling on the fully observed dataset (three to five large clusters with proportions over 0.1).

Among these extreme profiles, the two largest profiles are similar to the \textit{liberal} profile and the \textit{conservative} profile we have in modeling with the fully observed dataset. Table \ref{liberal-profile-MAR} shows the comparison of the \textit{liberal} profile obtained from the original dataset and that from the MAR scenario, focusing on the variables subjected to missing values. Table \ref{conservative-profile-MAR} reports a similar comparison for the \textit{conservative} profiles of the two scenarios. The profiles obtained from the data under the MAR scenario share the same modal responses and have similar distributions of answers to different questions as the profiles from the original dataset. They also have similar size to their respective counterparts in the analysis of the fully observed dataset. The analyses of cohesion ratios and disagreement scores of these clusters yield similar results to what we discussed previously for the fully observed dataset.

\begin{table}[t!]
\centering
\footnotesize
\caption{Comparison of the \textit{liberal} profiles of the variables subjected to missing values between the fully observed scenario and the MAR scenario.} 
\begin{tabular}{lrrr|rrr}
  \toprule
\multicolumn{1}{c}{}  &   \multicolumn{3}{c}{without missing data} & \multicolumn{3}{c}{MAR 30\%}\\
\multicolumn{1}{c}{}  &   \multicolumn{3}{c}{proportion 28.49\%} & \multicolumn{3}{c}{proportion 29.08\%}\\
\hline
& level 1 & level 2 & level 3 & level 1 & level 2 & level 3 \\ 
\hline
V162140: Favor/oppose tax on millionaires & 0.99 & 0.00 & 0.01 & 0.98 & 0.01 & 0.01 \\ 
  V162158: How likely immigrants take away jobs & 0.00 & 0.37 & 0.62 & 0.00 & 0.41 & 0.59 \\ 
  V162179: Should marijuana be legal & 0.80 & 0.04 & 0.16 & 0.81 & 0.05 & 0.15 \\ 
  V162207: World changing and we should adjust & 0.75 & 0.07 & 0.18 & 0.76 & 0.07 & 0.16 \\ 
  V162214: Blacks must try harder to get ahead & 0.00 & 0.03 & 0.97 & 0.01 & 0.02 & 0.97 \\ 
  V162269: U.S.’s culture is harmed by immigrants & 0.00 & 0.00 & 1.00 & 0.00 & 0.00 & 0.99 \\ 
  V162123: Better if rest of world more like U.S. & 0.05 & 0.20 & 0.76 & 0.05 & 0.18 & 0.76 \\ 
  V162148: Favor/oppose reduce income inequality & 0.82 & 0.03 & 0.15 & 0.82 & 0.03 & 0.15 \\ 
  V162170: Country needs strong leader & 0.03 & 0.13 & 0.84 & 0.03 & 0.15 & 0.83 \\ 
  V162192: Should the minimum wage be raised & 0.97 & 0.03 & 0.00 & 0.97 & 0.03 & 0.00 \\ 
  V162209: More tolerant of other moral standards & 0.95 & 0.03 & 0.01 & 0.96 & 0.02 & 0.02 \\ 
  V162246: Fewer problems if treated people fairly & 0.96 & 0.03 & 0.01 & 0.96 & 0.02 & 0.02 \\ 
  \bottomrule
\end{tabular}
\label{liberal-profile-MAR}
\end{table}

\begin{table}[t!]
\centering
\footnotesize
\caption{Comparison of the \textit{conservative} profiles of the variables subjected to missing values between the fully observed scenario and the MAR scenario.} 
\begin{tabular}{lrrr|rrr}
  \toprule
\multicolumn{1}{c}{}  &   \multicolumn{3}{c}{without missing data} & \multicolumn{3}{c}{MAR 30\%}\\
\multicolumn{1}{c}{}  &   \multicolumn{3}{c}{proportion 15.77\%} & \multicolumn{3}{c}{proportion 16.69\%}\\
\hline
& level 1 & level 2 & level 3 & level 1 & level 2 & level 3 \\ 
\hline
V162140: Favor/oppose tax on millionaires & 0.14 & 0.52 & 0.34 & 0.20 & 0.44 & 0.35 \\ 
  V162158: How likely immigrants take away jobs & 0.34 & 0.63 & 0.02 & 0.35 & 0.63 & 0.02 \\ 
  V162179: Should marijuana be legal & 0.17 & 0.64 & 0.20 & 0.27 & 0.52 & 0.21 \\ 
  V162207: World changing and we should adjust & 0.06 & 0.03 & 0.91 & 0.07 & 0.04 & 0.90 \\ 
  V162214: Blacks must try harder to get ahead & 0.84 & 0.13 & 0.04 & 0.88 & 0.08 & 0.04 \\ 
  V162269: U.S.’s culture is harmed by immigrants & 0.40 & 0.31 & 0.28 & 0.50 & 0.28 & 0.22 \\ 
  V162123: Better if rest of world more like U.S. & 0.53 & 0.38 & 0.09 & 0.50 & 0.42 & 0.08 \\ 
  V162148: Favor/oppose reduce income inequality & 0.01 & 0.95 & 0.04 & 0.02 & 0.90 & 0.09 \\ 
  V162170: Country needs strong leader & 0.91 & 0.07 & 0.02 & 0.92 & 0.07 & 0.02 \\ 
  V162192: Should the minimum wage be raised & 0.07 & 0.71 & 0.23 & 0.13 & 0.70 & 0.17 \\ 
  V162209: More tolerant of other moral standards & 0.19 & 0.19 & 0.61 & 0.20 & 0.20 & 0.61 \\ 
  V162246: Fewer problems if treated people fairly & 0.20 & 0.27 & 0.53 & 0.18 & 0.31 & 0.51 \\ 
  \bottomrule
\end{tabular}
\label{conservative-profile-MAR}
\end{table}

The entire simulation process was repeated multiple times, each time generating a new missingness pattern and a new initialization for the MCMC with different random seeds. The number of moderate and large clusters varies from three to seven dominant profiles, among which profiles similar to our \textit{conservative} and \textit{liberal} profiles of the fully observed dataset always appear consistently. Therefore, the HDPMPM model provides an alternative approach for mixture modeling that also accommodates the handling of missing values when present.

\subsection{MCAR scenario}
Since researchers sometimes have to analyze data under MCAR, this section demonstrates the performance of the HDPMPM model in this scenario. To incorporate MCAR, missing values are set independently for the twelve variables at a 30\% missingness rate.

There are six clusters with expected population proportions over 0.1. Among these extreme profiles, the two largest profiles are similar to the \textit{liberal} profile and the \textit{conservative} profile we have in modeling with the fully observed dataset. The other profiles show milder responses to different political ideas. 

Parallel to the analyses done for the MAR scenario, Table \ref{liberal-profile-MCAR} shows the comparison of the \textit{liberal} profile obtained from the original dataset and that from the MCAR scenario. Similarly, Table \ref{conservative-profile-MCAR} reports a comparison of the \textit{conservative} profiles. Again, the two extreme profiles from the survey under MCAR scenario closely match those of the original dataset in modal responses, answer distributions, and size relative to their fully observed counterparts. 

\begin{table}[t!]
\centering
\footnotesize
\caption{Comparison of the \textit{liberal} profiles of the variables subjected to missing values between the fully observed scenario and the MCAR scenario.} 
\begin{tabular}{lrrr|rrr}
  \toprule
\multicolumn{1}{c}{}  &   \multicolumn{3}{c}{without missing data} & \multicolumn{3}{c}{MCAR 30\%}\\
\multicolumn{1}{c}{}  &   \multicolumn{3}{c}{proportion 28.49\%} & \multicolumn{3}{c}{proportion 28.69\%}\\
\hline
& level 1 & level 2 & level 3 & level 1 & level 2 & level 3 \\ 
\hline
V162140: Favor/oppose tax on millionaires & 0.99 & 0.00 & 0.01 & 0.99 & 0.00 & 0.01 \\ 
  V162158: How likely immigrants take away jobs & 0.00 & 0.37 & 0.62 & 0.00 & 0.38 & 0.61 \\ 
  V162179: Should marijuana be legal & 0.80 & 0.04 & 0.16 & 0.80 & 0.04 & 0.15 \\ 
  V162207: World changing and we should adjust & 0.75 & 0.07 & 0.18 & 0.76 & 0.06 & 0.18 \\ 
  V162214: Blacks must try harder to get ahead & 0.00 & 0.03 & 0.97 & 0.01 & 0.01 & 0.98 \\ 
  V162269: U.S.’s culture is harmed by immigrants & 0.00 & 0.00 & 1.00 & 0.00 & 0.00 & 0.99 \\ 
  V162123: Better if rest of world more like U.S. & 0.05 & 0.20 & 0.76 & 0.04 & 0.19 & 0.77 \\ 
  V162148: Favor/oppose reduce income inequality & 0.82 & 0.03 & 0.15 & 0.82 & 0.03 & 0.14 \\ 
  V162170: Country needs strong leader & 0.03 & 0.13 & 0.84 & 0.02 & 0.13 & 0.85 \\ 
  V162192: Should the minimum wage be raised & 0.97 & 0.03 & 0.00 & 0.98 & 0.02 & 0.00 \\ 
  V162209: More tolerant of other moral standards & 0.95 & 0.03 & 0.01 & 0.95 & 0.04 & 0.01 \\ 
  V162246: Fewer problems if treated people fairly & 0.96 & 0.03 & 0.01 & 0.97 & 0.02 & 0.01 \\ 
  \bottomrule
\end{tabular}
\label{liberal-profile-MCAR}
\end{table}

\begin{table}[t!]
\centering
\footnotesize
\caption{Comparison of the \textit{conservative} profiles of the variables subjected to missing values between the fully observed scenario and the MCAR scenario.} 
\begin{tabular}{lrrr|rrr}
  \toprule
\multicolumn{1}{c}{}  &   \multicolumn{3}{c}{without missing data} & \multicolumn{3}{c}{MCAR 30\%}\\
\multicolumn{1}{c}{}  &   \multicolumn{3}{c}{proportion 15.77\%} & \multicolumn{3}{c}{proportion 13.05\%}\\
\hline
& level 1 & level 2 & level 3 & level 1 & level 2 & level 3 \\ 
\hline
V162140: Favor/oppose tax on millionaires & 0.14 & 0.52 & 0.34 & 0.14 & 0.50 & 0.36 \\ 
  V162158: How likely immigrants take away jobs & 0.34 & 0.63 & 0.02 & 0.33 & 0.65 & 0.02 \\ 
  V162179: Should marijuana be legal & 0.17 & 0.64 & 0.20 & 0.19 & 0.60 & 0.20 \\ 
  V162207: World changing and we should adjust & 0.06 & 0.03 & 0.91 & 0.03 & 0.04 & 0.92 \\ 
  V162214: Blacks must try harder to get ahead & 0.84 & 0.13 & 0.04 & 0.83 & 0.14 & 0.02 \\ 
  V162269: U.S.’s culture is harmed by immigrants & 0.40 & 0.31 & 0.28 & 0.41 & 0.34 & 0.26 \\ 
  V162123: Better if rest of world more like U.S. & 0.53 & 0.38 & 0.09 & 0.54 & 0.36 & 0.10 \\ 
  V162148: Favor/oppose reduce income inequality & 0.01 & 0.95 & 0.04 & 0.01 & 0.95 & 0.04 \\ 
  V162170: Country needs strong leader & 0.91 & 0.07 & 0.02 & 0.90 & 0.07 & 0.03 \\ 
  V162192: Should the minimum wage be raised & 0.07 & 0.71 & 0.23 & 0.07 & 0.71 & 0.22 \\ 
  V162209: More tolerant of other moral standards & 0.19 & 0.19 & 0.61 & 0.17 & 0.20 & 0.63 \\ 
  V162246: Fewer problems if treated people fairly & 0.20 & 0.27 & 0.53 & 0.20 & 0.31 & 0.49 \\ 
  \bottomrule
\end{tabular}
\label{conservative-profile-MCAR}
\end{table}

The simulation under the MCAR scenario was repeated multiple times with new missingness patterns and different initializations for the MCMC. The number of moderate and large clusters varies from three to seven dominant profiles, among which profiles similar to our \textit{conservative} and \textit{liberal} profiles of the fully observed dataset always appear consistently. Similar to what we have in the MAR scenario, the analyses of cohesion ratios and disagreement scores of these clusters yield similar results to what we obtained previously for the fully observed dataset.

\section{Discussion} \label{section_conclusion} 

In this article, we consider a practical problem in analyzing survey data consisting of multivariate categorical responses. Treating responses from each person as a group of variables, this paper proposes the HDPMPM model, which helps capture the latent structure in the form of the mixture of products of multinomial distributions while allowing the mixture components to be shared across different individuals. This mixed membership method uses HDP as the prior distribution for the mixture components. The Gibbs sampling scheme for sampling from the posterior distributions of all model parameters using the truncated stick-breaking representation of the DP is presented. 

An application of this model is demonstrated by extracting latent political ideologies in our population dataset obtained from the 2016 ANES survey data. By analyzing dominant mixture components, we successfully extract meaningful political profiles, which we call the \textit{liberal} profile and the \textit{conservative} profile. We then justify these labels using posterior expectations of mixture parameters, which generally agree with the common notions of American political ideology. Further analyses using other tools, such as the cohesion ratio or the disagreement score, are also possible given posterior samples of the HDPMPM model. We also extend the sampler for the model to incorporate handling missing data and successfully show that we can recover these two dominant political profiles accurately even when the dataset is under either the MAR scenario or the MCAR scenario. 


As with any other methods, it is recommended that analysts check the model assumptions carefully before using the HDPMPM model. Analysts also need to choose the model parameters carefully. As an example, for the maximum number of clusters, initial runs may be required to choose the adequate number of mixture components. As mentioned previously, models using a Dirichlet process prior may be inconsistent on the posterior number of clusters and, hence, analysts should not rely solely on this method in analyzing the exact number of mixture profiles. Moreover, DP mixture models may generate small clusters that do not reflect the underlying data generating process. Finally, there are many alternative models for analyzing this type of dataset. Future work may focus on comparing the inference obtained from alternative methods with that from the HDPMPM model. Applications of the HDPMPM model on datasets from other domains, such as healthcare, are also possible.

\section{Supplementary Material}

The supplementary material contains the derivation of the proposed Gibbs sampling scheme for posterior samples of parameters in the HDPMPM model, the data dictionary of our dataset with the exact wording of each survey item according to the 2016 ANES survey, and the logistic regressions to create dataset under the MAR scenario.  

\bigskip
{\raggedright
\textbf{Acknowledgment}

The author expresses sincere gratitude to Dr. Olanrewaju Akande, who served as the project advisor for this study. His insightful guidance and advice significantly shaped the direction of this work. The author also acknowledges Duke University for providing access to essential resources that supported the early phases of this research. 
}

\bibliographystyle{agsm}
\bibliography{references}

@book{littlerubin,
         author = {Little, {R. J. A.} and Rubin, D. B.},
         title = {Statistical Analysis with Missing Data: Third Edition},
         Year = 2019,
         Publisher = {New York: John Wiley \& Sons}
     }

@article{DPMPM_Kenough,
author = {Dunson, David and Xing, Chuanhua},
year = {2012},
month = {01},
pages = {1042-1051},
title = {Nonparametric Bayes Modeling of Multivariate Categorical Data},
volume = {104},
journal = {Journal of the American Statistical Association},
doi = {10.1198/jasa.2009.tm08439}
}

@book{rubin:1987,
  author = {Rubin, D. B.},
  pages = 258,
  publisher = {Wiley},
  title = {Multiple Imputation for Nonresponse in Surveys},
  year = 1987
}

@article{si:reiter:13,
author = {Si, Y. and Reiter, J. P.},
title = {Nonparametric {B}ayesian multiple imputation for incomplete categorical variables in large-scale assessment surveys},
journal ={Journal of Educational and Behavioral Statistics}, 
volume = 38,
year = 2013,
pages = {499--521}}

@article{manrique:reiter:sm,
author = {Manrique-{V}allier, D. and Reiter, J. P.},
Journal = {Survey Methodology},
Year = 2014,
title = {Bayesian multiple imputation for large-scale categorical data with structural zeros},
Volume =  40,
Pages = {125--134}}

@Misc{ANES2016,
author={{American National Election Studies}},
title={{ANES 2016 Time Series Study}},
year={2017},
publisher={Inter-university Consortium for Political and Social Research [distributor]},
doi={10.3886/ICPSR36824.v2},
url={https://doi.org/10.3886/ICPSR36824.v2}
}

@article{teh:2006,
author = {Yee Whye Teh and Michael I Jordan and Matthew J Beal and David M Blei},
title = {Hierarchical Dirichlet Processes},
journal = {Journal of the American Statistical Association},
volume = {101},
number = {476},
pages = {1566-1581},
year  = {2006},
publisher = {Taylor & Francis},
doi = {10.1198/016214506000000302},
eprint = { https://doi.org/10.1198/016214506000000302}
}

@inproceedings{ren:2008,
author = {Ren, Lu and Dunson, David B. and Carin, Lawrence},
title = {The dynamic hierarchical Dirichlet process},
year = {2008},
isbn = {9781605582054},
publisher = {Association for Computing Machinery},
address = {New York, NY, USA},
url = {https://doi.org/10.1145/1390156.1390260},
doi = {10.1145/1390156.1390260},
booktitle = {Proceedings of the 25th International Conference on Machine Learning},
pages = {824–831},
numpages = {8},
location = {Helsinki, Finland},
series = {ICML '08}
}

@article{woodbury:1978,
title = {Mathematical typology: A grade of membership technique for obtaining disease definition},
journal = {Computers and Biomedical Research},
volume = {11},
number = {3},
pages = {277-298},
year = {1978},
issn = {0010-4809},
doi = {https://doi.org/10.1016/0010-4809(78)90012-5},
author = {Max A. Woodbury and Jonathan Clive and Arthur Garson}
}

@article{Erosheva:2007,
 ISSN = {19326157},
 author = {Elena A. Erosheva and Stephen E. Fienberg and Cyrille Joutard},
 journal = {The Annals of Applied Statistics},
 number = {2},
 pages = {502--537},
 publisher = {Institute of Mathematical Statistics},
 title = {Describing Disability through Individual-Level Mixture Models for Multivariate Binary Data},
 volume = {1},
 year = {2007}
}

@inproceedings{gross:2012,
  author    = {Justin H. Gross and Daniel Manrique-Vallier},
  title     = {A Mixed-Membership Approach to the Assessment of Political Ideology from Survey Responses},
  booktitle = {Individual Presentation, Society for Political Methodology, 29th Annual Summer Meeting},
  address   = {Chapel Hill, NC},
  publisher = {Citeseer},
  year      = {2012}
}

@article{feldman:1988,
  title={Structure and Consistency in Public Opinion: the Role of Core Beliefs and Values},
  author={S. Feldman},
  journal={American Journal of Political Science},
  year={1988},
  volume={32},
  pages={416}
}

@book{maddox:1984,
  title={Beyond liberal and conservative: Reassessing the political spectrum},
  author={Maddox, William S and Lilie, Stuart A},
  year={1984},
  publisher={Cato Institute}
}

@article{feldman:2009,
author = {Feldman, Stanley and Johnston, Christopher},
year = {2009},
month = {01},
pages = {},
title = {Understanding the Determinants of Political Ideology: Implications of Structural Complexity},
volume = {35},
journal = {Political Psychology},
doi = {10.1111/pops.12055}
}

@article{seth94,
author = {Sethuraman, J.},
Year = 1994,
title = {A constructive definition of {D}irichlet priors},
Journal = {Statistica Sinica},
Volume = 4,
Pages = {639--650}
}

@article{Ishwaran2001,
author = {Hemant Ishwaran and Lancelot F James},
title = {Gibbs Sampling Methods for Stick-Breaking Priors},
journal = {Journal of the American Statistical Association},
volume = {96},
number = {453},
pages = {161-173},
year  = {2001},
publisher = {Taylor & Francis},
doi = {10.1198/016214501750332758},
eprint = { 
        https://doi.org/10.1198/016214501750332758
}
}

@article{Raftery2007,
author = {Raftery, Adrian and Newton, Michael and Satagopan, Jaya and Krivitsky, Pavel},
year = {2007},
month = {01},
pages = {},
title = {Estimating the integrated likelihood via posterior simulation using the harmonic mean identity},
volume = {8},
journal = {Bayesian statistics}
}

@article{dunsonxing,
author = {Dunson, D. B. and Xing, C.},
Year = 2009,
title = {Nonparametric {B}ayes modeling of multivariate categorical data},
Journal = {Journal of the American Statistical Association},
Volume = 104,
Pages = {1042--1051}}

@book{americanvoter,
  author = {Campbell, Angus and Converse, Philip E. and Miller, Warren E. and Stokes, Donald E.},
  publisher = {University of Chicago Press},
  title = {The American Voter},
  year = 1960
}

@article{Converse2006,
author = { Philip E.   Converse },
title = {The nature of belief systems in mass publics (1964)},
journal = {Critical Review},
volume = {18},
number = {1-3},
pages = {1-74},
year  = {2006},
publisher = {Routledge},
doi = {10.1080/08913810608443650},
eprint = { 
        https://doi.org/10.1080/08913810608443650
}
}

@book{zaller1992, 
place={Cambridge}, 
series={Cambridge Studies in Public Opinion and Political Psychology}, 
title={The Nature and Origins of Mass Opinion}, DOI={10.1017/CBO9780511818691}, 
publisher={Cambridge University Press}, 
author={Zaller, John R.}, 
year={1992}, 
collection={Cambridge Studies in Public Opinion and Political Psychology}}

@article{goodman1974,
    author = {Goodman, Leo A.},
    title = "{Exploratory latent structure analysis using both identifiable and unidentifiable models}",
    journal = {Biometrika},
    volume = {61},
    number = {2},
    pages = {215-231},
    year = {1974},
    month = {08},
    issn = {0006-3444},
    doi = {10.1093/biomet/61.2.215},
    eprint = {https://academic.oup.com/biomet/article-pdf/61/2/215/745752/61-2-215.pdf},
}

@article{Taylor1983,
author = {Taylor, Marylee C.},
title = {The Black-and-White Model of Attitude Stability: A Latent Class Examination of Opinion and Nonopinion in the American Public},
journal = {American Journal of Sociology},
volume = {89},
number = {2},
pages = {373-401},
year = {1983},
doi = {10.1086/227870},
eprint = { 
        https://doi.org/10.1086/227870
    
}
}

@techreport{pew2011,
  title = {Beyond Red vs. Blue: The Political Typology},
  author = {Andrew Kohut and Carroll Doherty and Michael Dimock and Scott Keeter},
  institution = {Pew Research Center},
  year = {2011},
  month = May,
  day = 4,
  address = {{Washington, D.C.}},
  url ={https://www.pewresearch.org/politics/2011/05/04/beyond-red-vs-blue-the-political-typology/},
  language = {en-US}
}

@article{Miller2014,
  author  = {Jeffrey W. Miller and Matthew T. Harrison},
  title   = {Inconsistency of Pitman-Yor Process Mixtures for the Number of Components},
  journal = {Journal of Machine Learning Research},
  year    = {2014},
  volume  = {15},
  number  = {96},
  pages   = {3333-3370},
  url     = {http://jmlr.org/papers/v15/miller14a.html}
}

@inproceedings{Miller2013,
 author = {Miller, Jeffrey W and Harrison, Matthew T},
 booktitle = {Advances in Neural Information Processing Systems},
 editor = {C. J. C. Burges and L. Bottou and M. Welling and Z. Ghahramani and K. Q. Weinberger},
 pages = {},
 publisher = {Curran Associates, Inc.},
 title = {A simple example of Dirichlet process mixture inconsistency for the number of components},
 volume = {26},
 year = {2013}
}

@article{Green2001,
author = {Green, Peter J. and Richardson, Sylvia},
title = {Modelling Heterogeneity With and Without the Dirichlet Process},
journal = {Scandinavian Journal of Statistics},
volume = {28},
number = {2},
pages = {355-375},
doi = {https://doi.org/10.1111/1467-9469.00242},
eprint = {https://onlinelibrary.wiley.com/doi/pdf/10.1111/1467-9469.00242},
year = {2001}
}

@misc{guha2019posterior,
      title={On posterior contraction of parameters and interpretability in Bayesian mixture modeling}, 
      author={Aritra Guha and Nhat Ho and XuanLong Nguyen},
      year={2019},
      eprint={1901.05078},
      archivePrefix={arXiv},
      primaryClass={math.ST}
}

@article{spearman04,
  author = {Spearman, Charles},
  journal = {American Journal of Psychology},
  key = {Spearman},
  pages = {201--293},
  title = {``{General intelligence},'' objectively determined and measured},
  volume = 15,
  year = 1904
}

@article{blackwell73,
 ISSN = {00905364, 21688966},
 URL = {http://www.jstor.org/stable/2958021},
 author = {David Blackwell},
 journal = {The Annals of Statistics},
 number = {2},
 pages = {356--358},
 publisher = {Institute of Mathematical Statistics},
 title = {Discreteness of Ferguson Selections},
 urldate = {2024-12-08},
 volume = {1},
 year = {1973}
}

@article{Hagenaars2003,
author = {Hagenaars, Jacques and Mccutcheon, Allan},
year = {2003},
month = {01},
pages = {},
title = {Applied Latent Class Analysis Models},
volume = {28},
journal = {Canadian Journal of Sociology / Cahiers canadiens de sociologie},
doi = {10.2307/3341848}
}

@article{Meyer2019UsingTD,
  title={Using the Dirichlet process to form clusters of people’s concerns in the context of future party identification},
  author={Patrick Meyer and Fenja M. Schophaus and Thomas Glassen and Jasmin Riedl and Julia M. Rohrer and Gert G. Wagner and Timo von Oertzen},
  journal={PLoS ONE},
  year={2019},
  volume={14}
}

@article{Leroux1992,
 ISSN = {00905364, 21688966},
 URL = {http://www.jstor.org/stable/2242015},
 author = {Brian G. Leroux},
 journal = {The Annals of Statistics},
 number = {3},
 pages = {1350--1360},
 publisher = {Institute of Mathematical Statistics},
 title = {Consistent Estimation of a Mixing Distribution},
 urldate = {2024-12-09},
 volume = {20},
 year = {1992}
}

@article{Fleishman1986,
 ISSN = {0033362X, 15375331},
 URL = {http://www.jstor.org/stable/2748725},
 author = {John A. Fleishman},
 journal = {The Public Opinion Quarterly},
 number = {3},
 pages = {371--386},
 publisher = {[Oxford University Press, American Association for Public Opinion Research]},
 title = {Types of Political Attitude Structure: Results of a Cluster Analysis},
 urldate = {2024-12-20},
 volume = {50},
 year = {1986}
}

@inbook{Silva2018,
title = "Public opinion surveys: a new scale",
author = "Guillem Rico and Eva Anduiza and Silva, {Bruno Castanho} and Ioannis Andreadis and Neboj{\v s}a Blanu{\v s}a and Corti, {Yazmin Morlet} and Gisela Delfino and Ruth-Lovell, {Saskia P.} and Bram Spruyt and Marco Steenbergen and Levente Littvay",
note = "Publisher Copyright: {\textcopyright} 2019 selection and editorial matter, Kirk A. Hawkins, Ryan E. Carlin, Levente Littvay, and Crist{\'o}bal Rovira Kaltwasser.",
year = "2018",
month = jan,
day = "1",
doi = "10.4324/9781315196923-8",
language = "English",
isbn = "9781138716513",
pages = "150--178",
booktitle = "The Ideational Approach to Populism",
publisher = "Taylor and Francis AS",
}

@article{treier2009,
    author = {Treier, Shawn and Hillygus, D. Sunshine},
    title = {The Nature of Political Ideology in the Contemporary Electorate},
    journal = {Public Opinion Quarterly},
    volume = {73},
    number = {4},
    pages = {679-703},
    year = {2009},
    month = {12},
    issn = {0033-362X},
    doi = {10.1093/poq/nfp067},
    url = {https://doi.org/10.1093/poq/nfp067},
    eprint = {https://academic.oup.com/poq/article-pdf/73/4/679/17163446/nfp067.pdf},
}

@article{Alvarez2017,
author = {Alvarez, R. Michael and Levin, Ines and N\'{u}\~{n}ez, Lucas},
title = {The Four Faces of Political Participation in Argentina: Using Latent Class Analysis to Study Political Behavior},
journal = {The Journal of Politics},
volume = {79},
number = {4},
pages = {1386-1402},
year = {2017},
doi = {10.1086/692786},

URL = { 
    
        https://doi.org/10.1086/692786
},
eprint = { 
    
        https://doi.org/10.1086/692786

}
}

@article{LAMERIS2018417,
title = {On the measurement of voter ideology},
journal = {European Journal of Political Economy},
volume = {55},
pages = {417-432},
year = {2018},
issn = {0176-2680},
doi = {https://doi.org/10.1016/j.ejpoleco.2018.03.003},
url = {https://www.sciencedirect.com/science/article/pii/S017626801730277X},
author = {Maite D. Laméris and Richard Jong-A-Pin and Harry Garretsen}
}

@article{Si_Reiter_Hillygus_2015, title={Semi-parametric Selection Models for Potentially Non-ignorable Attrition in Panel Studies with Refreshment Samples}, volume={23}, DOI={10.1093/pan/mpu009}, number={1}, journal={Political Analysis}, author={Si, Yajuan and Reiter, Jerome P. and Hillygus, D. Sunshine}, year={2015}, pages={92–112}}

@article{BREEN_2000, title={Why Is Support for Extreme Parties Underestimated by Surveys? A Latent Class Analysis}, volume={30}, DOI={10.1017/S0007123400230159}, number={2}, journal={British Journal of Political Science}, author={Breen, Richard}, year={2000}, pages={375–382}}

@article{MCCUTCHEON1985,
    author = {Mccutcheon, Allan L.},
    title = {A Latent Class Analysis of Tolerance for Nonconformity in the American Public},
    journal = {Public Opinion Quarterly},
    volume = {49},
    number = {4},
    pages = {474-488},
    year = {1985},
    month = {01},
    issn = {0033-362X},
    doi = {10.1086/268945},
    url = {https://doi.org/10.1086/268945},
    eprint = {https://academic.oup.com/poq/article-pdf/49/4/474/5292575/49-4-474.pdf},
}

@book{Airoldi2014,
author = {Airoldi, Edoardo M. and Blei, David and Erosheva, Elena A. and Fienberg, Stephen E.},
title = {Handbook of Mixed Membership Models and Their Applications},
year = {2014},
isbn = {1466504080},
publisher = {Chapman \& Hall/CRC},
edition = {1st}
}

\end{document}